\documentclass[10pt,floatfix,twocolumn,nobalancelastpage]{revtex4-1}

\usepackage{graphicx}
\usepackage{float}
\usepackage{amsmath}
\usepackage{physics}
\usepackage{siunitx}
\usepackage{float}
\usepackage[caption=false]{subfig}

\begin{document}

\title{Inverse Design of Near Unity Efficiency Perfectly Vertical Grating Couplers}

\author{Andrew Michaels*}
\author{Eli Yablonovitch}

\affiliation{Department of Electrical Engineering and Computer Science, 253 Cory Hall, Berkeley CA, 94720 \\
*Corresponding author: amichaels@berkeley.edu}

\date{\today}

\begin{abstract}
		Efficient coupling between integrated optical waveguides and optical fibers is essential to the success of integrated photonics.  While many solutions exist, perfectly vertical grating couplers which scatter light out of a waveguide in the direction normal to the waveguide's top surface are an ideal candidate due to their potential to reduce packaging complexity.  Designing such couplers with high efficiency, however, has proven difficult.  In this paper, we use electromagnetic inverse design techniques to optimize a high efficiency two-layer perfectly vertical silicon grating coupler.  Our base design achieves a chip-to-fiber coupling efficiency of over 99\% (-0.04 dB) at 1550 nm.  Using this base design, we apply subsequent constrained optimizations to achieve vertical couplers with over 96\% efficiency which are fabricable using a \SI{65}{\nano\meter} process.
\end{abstract}

\maketitle

\section{Introduction}

Optical couplers are an essential component in integrated photonic systems.   Grating couplers in particular present a number of advantages over alternative coupling methods\cite{mekis_grating-coupler-enabled_2011} by providing a flexible means of interfacing high-index-contrast integrated optical devices with the outside world. In general, grating couplers consist of a waveguide with either a partially or fully etched periodic corrugation on its top surface which causes light to scatter out of the waveguide at a desired angle. Of particular interest are grating couplers that couple light normally to the waveguides top surface, which is beneficial in terms of packaging complexity\cite{chen_fabrication-tolerant_2008}.  Designing perfectly vertical grating couplers with high efficiencies, however, turns out to be a difficult task; in addition to the challenges associated with mode matching and directionality which are relevant to all grating couplers, perfectly vertical grating couplers must contend with high second order Bragg reflections that arise as a result of the required periodicity of the grating. Tackling these three challenges is essential to realizing high-efficiency perfectly vertical gratings.

The problem of mode matching arises out of the need to couple light into a desired output mode (like that of an optical fiber).  Light scattered from a uniform grating whose corrugation has a fixed period and duty factor will have an intensity profile which decays exponentially along the length of the grating coupler\cite{tamir_analysis_1977}.  For the purpose of coupling to an optical fiber, this exponential intensity profile is not well matched to a fiber's fundamental mode and results in inefficient coupling.  We can rectify this problem by introducing a chirp to the grating duty factor and period.  By varying the duty factor, the strength with which light is scattered from the grating can be controlled, and thus a desired intensity (and phase) profile can be achieved\cite{taillaert_compact_2004}. 

In addition to mode matching, a grating coupler's efficiency is also limited by its directionality, i.e. the fraction of light that is scattered in the desired direction.  For a typical partially etched grating coupler, the directionality can be improved by properly setting the relative depth of the etch and the thickness of the remaining unetched waveguide below it \cite{vermeulen_high-efficiency_2010}.  Alternatively, this situation can be improved by adding material layers below the grating coupler in order to reflect light that is scattered below the grating.  In silicon, for example, an optimized silicon-on-insulator (SOI) grating coupler which uses the silicon substrate as a mirror can achieve total coupling efficiencies of over 60\% \cite{taillaert_grating_2006}.  Adding additional SOI layers below the grating coupler that act as a distributed Bragg reflector can further increase the directionality and total coupling efficiency to over 90\% \cite{taillaert_compact_2004}. While effective in increasing the efficiency, this approach of adding mirrors below the grating may not be acceptable in all situations (e.g. in a multilayer optical substrate in which vertical real estate is costly).  

Assuming we can achieve high directionality and mode-matching efficiency, the efficiency of a perfectly vertical grating coupler is still limited as a result of second order Bragg reflections.  The reason for this is easily understood by noting that the relationship between wavelength and the period of a grating coupler reduces to $\lambda/n_g = \Lambda$ where $\lambda$ is the wavelength, $n_g$ is the effective index of the grating, and $\Lambda$ is the period of the grating.  This is identical to the Bragg condition for second order reflections, and thus gratings which couple vertically are also susceptible to significant back reflections.  In general, this problem is simply avoided by choosing a grating period which couples light out at a small angle from normal \cite{taillaert_compact_2004}.  A number of solutions which do not sacrifice perfect vertical coupling have been proposed, such as introducing a diagonal etch \cite{wang_embedded_2005}, adding a reflecting notch at the input of the grating coupler \cite{roelkens_high_2007}, or adding a chirped section to the beginning of the grating \cite{chen_fabrication-tolerant_2008}.  Despite these efforts, chip-to-fiber efficiencies above 80\% have not to our knowledge been reported.

\begin{figure}[!htp]
		\centering
		\includegraphics[width=0.9\columnwidth]{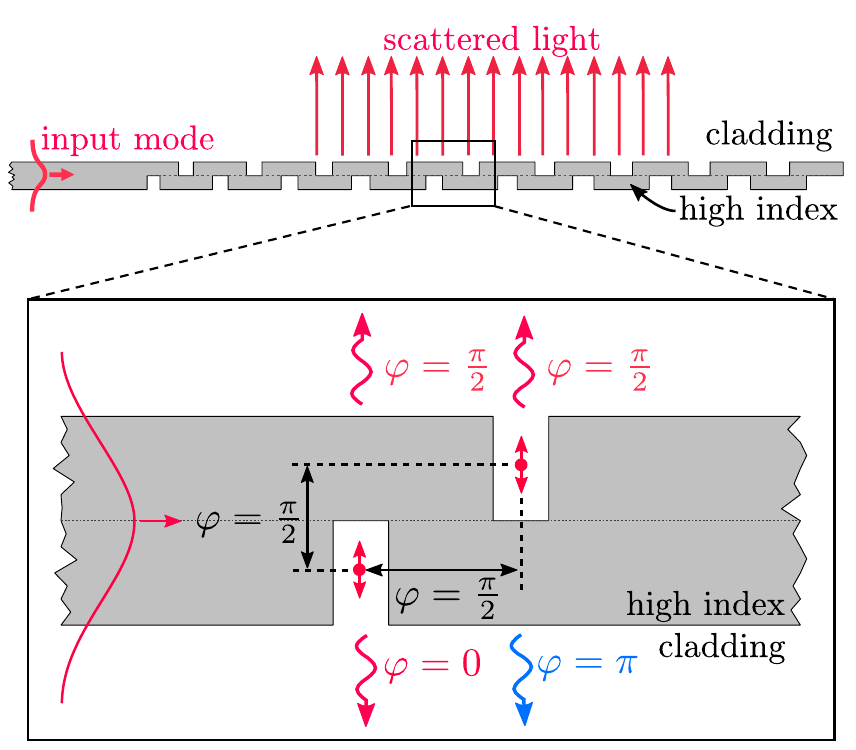}
		\caption{Description of the operation of two-layer grating couplers.  The two-layer grating resembles two fully-etched grating couplers which have been stacked one on top of the other with the top layer shifted forward by a small amount relative to the bottom layer. The thicknesses of the two layers as well as the distance between two subsequent gaps in the bottom and top layer are chosen such that light scattered in the upwards direction by each layer is in phase and interferes constructively while light scattered in the downwards direction by the two layers is $\pi$ out of phase and interferes destructively.  This enables grating couplers to scatter close to 100\% of the input light in the upward direction.}
		\label{fig:two_layer_grating_principle}
\end{figure}

All three of these challenges, however, can be eliminated using a new type of grating coupler which has recently received attention due to its inherent high directionality \cite{notaros_ultra-efficient_2016,dai_highly_2015}.  These gratings consist of two partially- or fully-etched grating couplers which have been stacked on top of one another.  When designed properly, these ``two-layer gratings'' act as a phased array of scatterers which couple light out of the waveguide with a high directionality.  The operating principle of these gratings is depicted in Fig. \ref{fig:two_layer_grating_principle}.  A guided wave propagating along the grating will encounter grooves in the top and bottom surfaces of the waveguide which couple some of the propagating power into free-space modes.  If the horizontal separation between the bottom grooves and top grooves is chosen such that the guided mode accumulates a phase shift of $\pi/2$ as it propagates from a groove in the bottom layer to its adjacent top layer groove and the thicknesses of the top and bottom layers and etch depths of the grating are chosen such that a wave scattered out of a bottom groove acquires an additional $\pi/2$ phase shift relative to a wave scattered out of a top groove, the resulting waves will interfere constructively in the upwards direction and destructively in the downwards direction. In addition to enabling a directionality of nearly 100\%, the relative shift between the top and bottom layers produces two sets of reflected guided waves within the grating which are $\pi$ out of phase.  These waves destructively interfere, resulting in an exceptionally low back reflection, even for perfectly vertical coupling.  Finally, the addition of a second layer of scatters increases the number of designable degrees of freedom which improves our ability to match to a desired mode.

These properties make two-layer grating couplers an extremely attractive solution for perfectly vertical optical coupling.  Our goal moving forward is thus to design a perfectly vertical two layer grating which achieves a high efficiency and is fabricable.  We achieve this using electromagnetic optimization techniques described in the next section.

\section{Grating Coupler Optimization}

The introduction of a second layer inherently increases the complexity of the grating coupler and hence the difficulty associated with designing an efficient structure. In order to handle this complexity, we have employed gradient-based shape optimization techniques with boundary smoothing \cite{michaels_gradient-based_2017} which are capable of rapidly optimizing electromagnetic structures with many degrees of freedom.  These methods are particularly well suited to grating couplers whose shapes are constrained in a well-defined way.

Successful optimization of a grating coupler is strongly contingent on three important choices: (1) the figure of merit that properly describes its efficiency, (2) the parameterization of the geometry, and (3) the initial choice for the geometry.  In general, we design grating couplers with the intention of coupling an input waveguide mode to a desired output waveguide (an optical fiber for example) or free-space mode.  As is such, the most appropriate figure of merit is the mode overlap between the fields scattered from the grating and a desired field profile.  This mode overlap describes the fraction of power in an electromagnetic field which can couple into a desired mode, which in the case of grating couplers is most commonly the approximately Gaussian field profile of an optical fiber.   Assuming that no light is reflected back towards the grating, the mode overlap expression used to calculate the grating coupler efficiency can be simplified to \cite{michaels_gradient-based_2017} 

\begin{equation}
		\eta =  \frac{1}{4 P_m P_\mathrm{src}}\left|\iint\limits_A d\mathbf{A} \cdot \mathbf{E} \times \mathbf{H}_m^*\right|^2 
		\label{eq:mode_overlap_simple}
\end{equation}

\noindent where $\mathbf{E}$ is the incident electric field, $\mathbf{H}_m$ is the magnetic fields of the desired mode profile, and $P_\mathrm{src}$ and $P_m$ are the total input source power to the system and the power carried in the desired mode, respectively.The integral is computed over a plane large enough to encompass the incident and desired fields.  For our purpose, $\mathbf{E}_m$ and $\mathbf{H}_m$ correspond to an electric and magnetic field with a Gaussian intensity profile and flat phase profile.

The optimization process consists of improving the grating coupler efficiency given by Eq. (\ref{eq:mode_overlap_simple}) by varying designable parameters which define the structure.  This necessitates that we choose a parameterization of our grating coupler.  A straight forward choice of parameterization is to make the width of each gap and tooth in the grating an independent parameter.  While this parameterization is very flexible, it is susceptible to producing strange or irregular structures.  The nature in which grating couplers leak power is relatively well behaved, and thus it is unlikely that such irregular structures are optimal for generating a beam which matches, for example, a smoothly varying Gaussian. 

\begin{figure}[!htp]
		\centering
		\includegraphics[width=1.0\columnwidth]{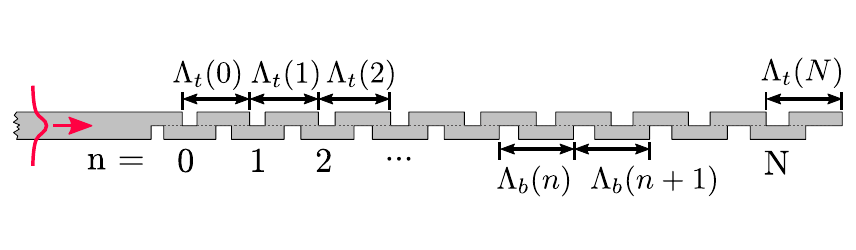}
		\caption{The grating coupler is parameterized in terms of local ``periods'' and duty factors which evolve along the length of the grating.  The distance between the onset of one gap and the next gap of the grating is the period $\Lambda$ while the duty factor describes the fraction of this period that is unetched.  Each period of the grating is assigned an index $n$ and the way in which the period and duty factor evolve along the grating is defined using a smooth function of this index $n$.}
		\label{fig:grating_indexing}
\end{figure}

Instead, it is advantageous to choose a parameterization which we can guarantee will generate a well behaved grating chirp.  To this end, a Fourier series parameterization of the grating period and duty factor is well suited.  To understand this parameterization, consider Fig. \ref{fig:grating_indexing}.  Each tooth and immediately preceding gap of the grating is assigned an (increasing) integer index.  The local period, labeled $\Lambda_t$ and $\Lambda_b$ (where the $t$ refers to the top layer and the $b$ refers to the bottom layer) in Fig. \ref{fig:grating_indexing}, are then defined as a function of this index.  In our optimizations, the functional form of the period is expressed as a Fourier series expansion given by

\begin{align}
		\Lambda(n) = a_0 &+ \sum\limits_{m=1}^{M} a_m \sin\left( \frac{\pi}{2} \frac{m}{N} n\right) \\\nonumber
						 &+  \sum\limits_{m=1}^{M} b_m \cos\left( \frac{\pi}{2} \frac{m}{N} n\right) 
		\label{eq:chirp_function}
\end{align}

\noindent where $n$ is the grating index, $M$ is the total number of Fourier terms, and $N$ is the number of periods in the grating.  In this expression, the coefficients $a_n$ and $b_n$ are the design parameters of the grating; by modifying their values, an arbitrary grating chirp can be realized (assuming enough Fourier terms are present).  In order to control how rapidly the dimensions of the grating can change along the grating's length, the number of Fourier terms can be reduced.  The duty factor, meanwhile, is expressed in an identical fashion.

A final requirement for the optimization of these couplers (and grating couplers in general) is the selection of a ``good'' starting structure.  The choice of a good starting point is particularly important for convex optimization techniques which make no guarantees about global optimality of the solution.  This is further complicated by the wave nature of electromagnetics which tends to introduce many local optima.  In the case of grating couplers, however, we can take advantage of our understanding of the physics of the problem in order to select a starting structure that will enable even a local optimization technique to find a highly efficient structure.  In particular, we know that the strength with which we scatter light must gradually increase from zero along the length of the grating in order to produce the desired Gaussian mode profile.  The primary way to accomplish this is with a duty factor which is initially very high (i.e. the grooves of the grating are very small) and slowly decreases over the length of the grating.  Based on this intuition, it makes sense to choose a starting structure with a high duty factor.  Given an initial duty factor, the period should be chosen to generate a beam that propagates at the correct angle in order to ensure that our starting design is as close as possible to a desirable solution.  In the case of the two-layer grating with a large duty factor, the period is given by (see Appendix \ref{sec:two_layer_equations} for derivation) 

\begin{equation}
		\Lambda = \frac{m \lambda }{(1-D)(n_{e2}+n_{e3}) + (2 D - 1) n_{e1} - n_{0}\sin\theta} \;\; .
		\label{eq:grating_period}
\end{equation}

\noindent where $\lambda$ is the free-space wavelength, $D$ is the duty factor, $n_{e1}$, $n_{e2}$, and $n_{e3}$ are the effective indices of a wave propagating through the different sections of the grating (depicted in Fig. \ref{fig:grating_parameter_diagram}), $n_{0}$ is the cladding index, $\theta$ is the angle relative to normal that the generated beam makes, and $m$ is the desired diffraction order (which in general is 1).  

With a starting geometry, grating parameterization, and an appropriate figure of merit selected, we may optimize our structure with relative confidence that a reasonable solution will be found. 

\begin{figure}[htb!]
		\centering
		\includegraphics[width=\columnwidth]{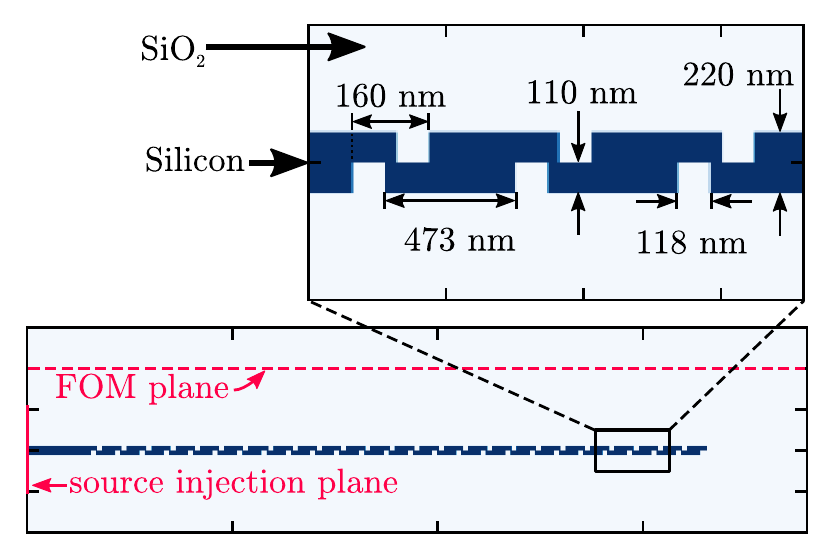}
		\caption{Plot of the initial starting geometry used in the optimization of a two layer SiO$_2$-clad silicon grating coupler.  A uniform grating with a duty factor of 80\% and a period of \SI{586}{\nano\meter} is chosen for both the top and bottom layers.  This choice of period and duty factor results in a nearly vertical beam with a high directionality.  }
		\label{fig:starting_geo}
\end{figure}

\begin{figure*}[htb!]
		\centering
		\includegraphics[width=\textwidth]{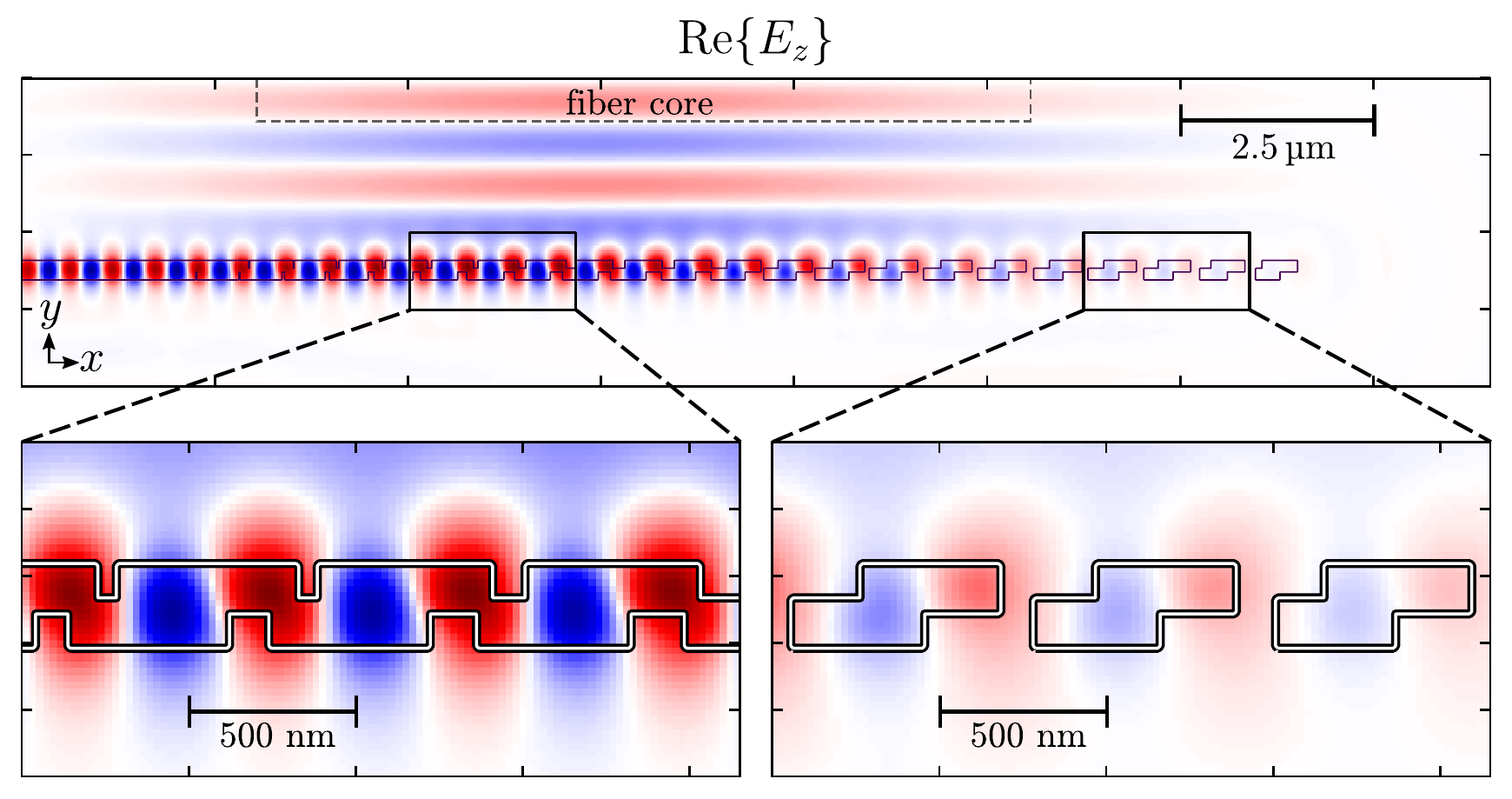}
		\caption{Optimization results for our perfectly vertical two layer grating coupler.  The real part of $E_z$, which has been overlayed with an outline of the optimized refractive index, shows perfectly vertical emission with an extremely high directionality and flat wavefronts.  This optimized structure is very well mode-matched to the mode of a \SI{10}{\micro\meter} mode field diameter single mode fiber located \SI{2}{\micro\meter} above the grating surface which is reflected by a chip-to-fiber efficiency of over 99\% ($>$-0.04 dB).}
		\label{fig:Silicon_2L_PV_results}
\end{figure*}

\section{Optimization Results}

For the purpose of these optimizations, we use a custom finite difference frequency domain (FDFD) solver to simulate Maxwell's equations.  Easy access to the internals of the simulation software simplifies the application of our optimization methods.  Because 1D grating couplers are significantly wider than they are thick, we make use of the effective index method to reduce the problem from three dimensions to two dimensions without a significant loss of accuracy, significantly speeding up the optimization process.

The starting structure for the optimization is shown in Fig. \ref{fig:starting_geo} and consists of a uniform two-layer silicon grating clad both on top and underneath with silicon dioxide.  We choose the layer thicknesses to be \SI{110}{\nano\meter} to produce the desired $\pi/2$ phase shift between the top and bottom layer.  Based on Eq. \ref{eq:grating_period}, a starting grating period of \SI{586}{\nano\meter}, a duty factor of 80\%, and a shift between the top and bottom layers of \SI{160}{\nano\meter} are chosen to ensure that the starting efficiency is reasonably high, which in this case was $\sim 48$\%.  The grating structure is excited by the fundamental mode of the waveguide shown at the left edge of Fig. \ref{fig:starting_geo} at a wavelength of \SI{1550}{\nano\meter}.  In the following optimizations, we only consider this single wavelength (although the same method could be applied to a broadband figure of merit if desired).

This initial structure is modified using our gradient-based optimization methods in order to maximize the efficiency with which the grating couples light into a \SI{10}{\micro\meter} mode field diameter Gaussian beam corresponding to the approximate mode of a single mode fiber which is situated \SI{2}{\micro\meter} above the grating and oriented normally to its top surface.  This efficiency is evaluated using Eq. \ref{eq:mode_overlap_simple} where $\mathbf{H}_m$ is the magnetic field of the desired Gaussian fiber mode. It is worth noting that while the modes of real single mode fibers may not be exactly Gaussian, our method can be used for any desired field profile (i.e. with non-Gaussian $\mathbf{H}_m$) as long as that field is known.

The optimization process is executed until the figure of merit changes by less than $10^{-5}$, which takes 67 iterations of the minimzation algorithm (requiring 166 simulations in total and about two hours of computation time on a single computer core).  The corresponding optimized structure and simulated electric field is displayed in Fig. \ref{fig:Silicon_2L_PV_results}.  As desired, light is coupled perfectly vertically and entirely in the upwards direction.  The wavefronts, furthermore, are very flat and well-behaved; the quality of this generated beam is reflected by an exceptionally high optimized chip-to-fiber coupling efficiency of 99.2\% (-0.034 dB).

\begin{figure}[!htb]
		\centering
		\includegraphics[width=\columnwidth]{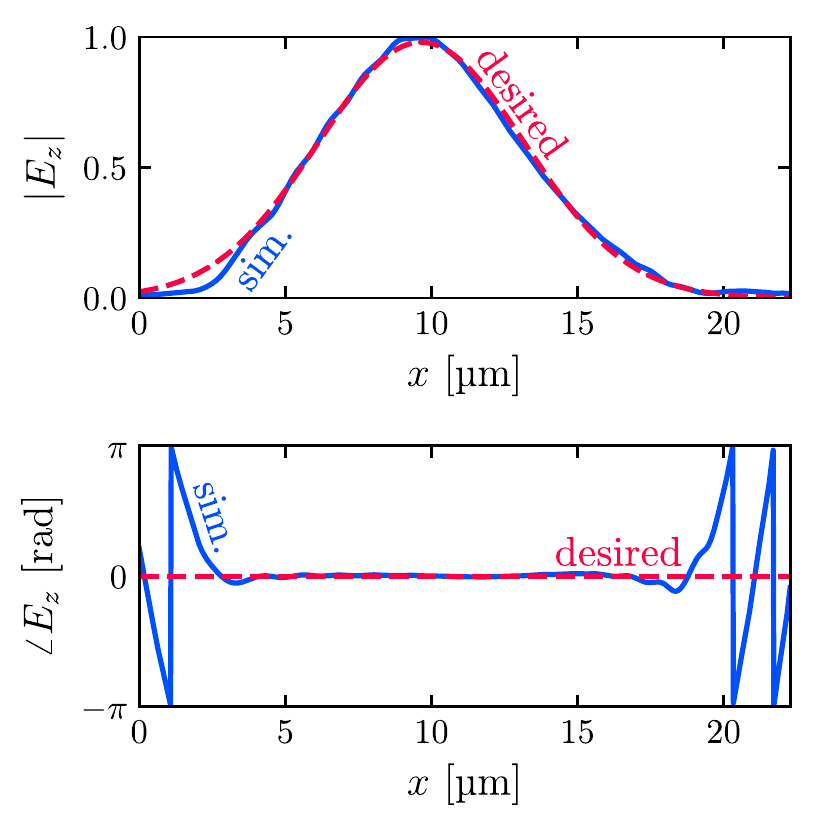}
		\caption{Plots comparing the simulated electric field amplitude (top) and phase (bottom) of the optimal grating coupler design to the desired amplitude and phase.  The visible deviation in the simulated phase is inconsequential as it occurs only when the field amplitude is very small.}
		\label{fig:mode_match}
\end{figure}

The mode matching capabilities of the optimized grating are more quantitatively demonstrated by the slices of $E_z$ shown in Fig. \ref{fig:mode_match}.  As is readily apparent, the magnitude of the simulated electric field very closely matches the desired Gaussian field profile, with the exception of some weak rippling.  We attribute this rippling to the inherently discrete nature of the grating coupler's rectangular scatters. However, as demonstrated by the calculated mode match of over 99\%, this has a minimal impact on the over all efficiency.  As with the electric field amplitude, the simulated phase is very flat over the majority of the beam's width, deviating only near the edges of the beam.  This deviation, however, does not lead to an appreciable decrease in efficiency since the field amplitude is nearly zero far away from the beam's center where the phase begins to fluctuate.  In addition to mode matching, the reflection of waveguide modes incident on the grating coupler had to be suppressed in order to achieve a very high efficiency.  This was indeed the case, and we calculated a back reflection of only \SI{-40.2}{\decibel} for our optimized design.  

\begin{figure}[htp!]
		\centering
		\includegraphics[width=0.9\columnwidth]{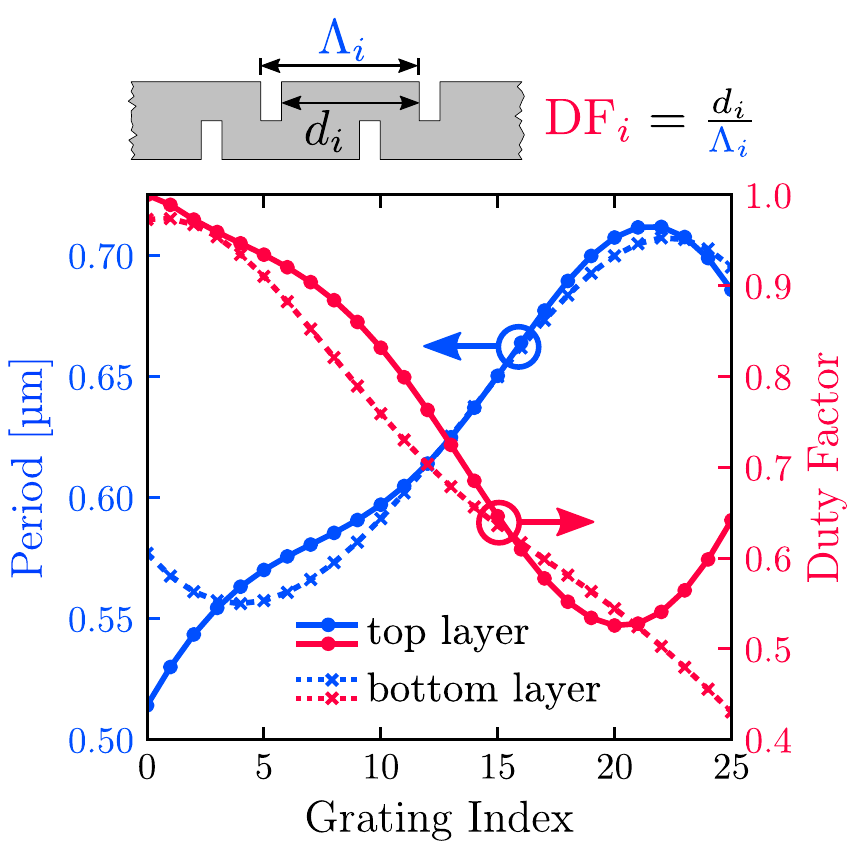}
		\caption{Plot of the chirp functions for the optimized two-layer grating.  The blue curves show the the period as a function of the position (index) along the grating.  The red curves, meanwhile, show the duty factor along the grating. In both sets of curves, the solid trace corresponds to the top layer while the dotted curve corresponds to the bottom layer of the grating.  }
		\label{fig:chirp_function}
\end{figure}

The unprecedented efficiency of this optimized grating is a direct consequence of the chirping of the grating period and duty factor visible in the insets of Fig. \ref{fig:Silicon_2L_PV_results} which is plotted as a function of grating index (position along the grating) in Fig. \ref{fig:chirp_function}.  The optimal chirp function for the duty factor (upon which the scattering strength of the grating depends most strongly) of the top layer roughly matches the behavior of the theoretical scattering parameter for an ideal grating coupler matched to a Gaussian field profile \cite{waldhausl_efficient_1997}.  Meanwhile, chirp function of the bottom layer's duty factor deviates from this ideal behavior, highlighting the strength of our optimization methods: its ability to design structures that would otherwise be difficult or impossible to design ``by hand.''

Fig. \ref{fig:chirp_function} also highlights the primary shortcoming of these optimization results: the optimal design contains duty factors that approach 100\%.  This means that the optimal structure contains features that are as small as a few nanometers wide and hence are not fabricable (at least in any practical setting).  Intuitively, however, if we are willing to sacrifice some amount of efficiency, we should be able to constrain the design such that non-fabricable features do not appear in the optimal design.  One way of doing this would be to simply impose a minimum feature size on the optimized design; this brute force approach, however, would inevitably leave us with a non-optimal design.  

Instead, it is desirable to introduce this minimum feature constraint directly into the optimization.  In order to accomplish this, we add an additional penalty term to our figure of merit.  Our original figure of merit is modified such that the efficiency of the grating is penalized when small features form. The new figure of merit is given by 

\begin{equation}
		F(\mathbf{p}) = \eta - f_\text{penalty}(\mathbf{p})
		\label{eq:constrained_FOM}
\end{equation}

\noindent where $F(\mathbf{p})$ is our new constrained figure of merit, $\mathbf{p}$ is the set of design variables (i.e. grating dimensions), $\eta$ is the mode match efficiency discussed previously, and $f_\text{penalty}(\mathbf{p})$ is a function of the design variables which penalizes the efficiency when the feature sizes of the grating are too small.  For this penalty function, we use an analytic approximation of a rect function which is positive when a feature in the grating is greater than zero and smaller than the specified minimum feature size and zero otherwise.  This function is computed for each gap width in the grating and summed up before subtracting from the efficiency.  Because this penalization process operates on each individual gap and tooth in the grating, we no longer use a Fourier series parameterization in subsequent optimizations, but instead opt for parameterization in which each gap and tooth dimension is a separate independent design variable.  Finally, in order to control the influence of this penalty function, we adjust its maximum value and the steepness of its edges.

\begin{figure}[!htp]
		\centering
		\includegraphics[width=\columnwidth]{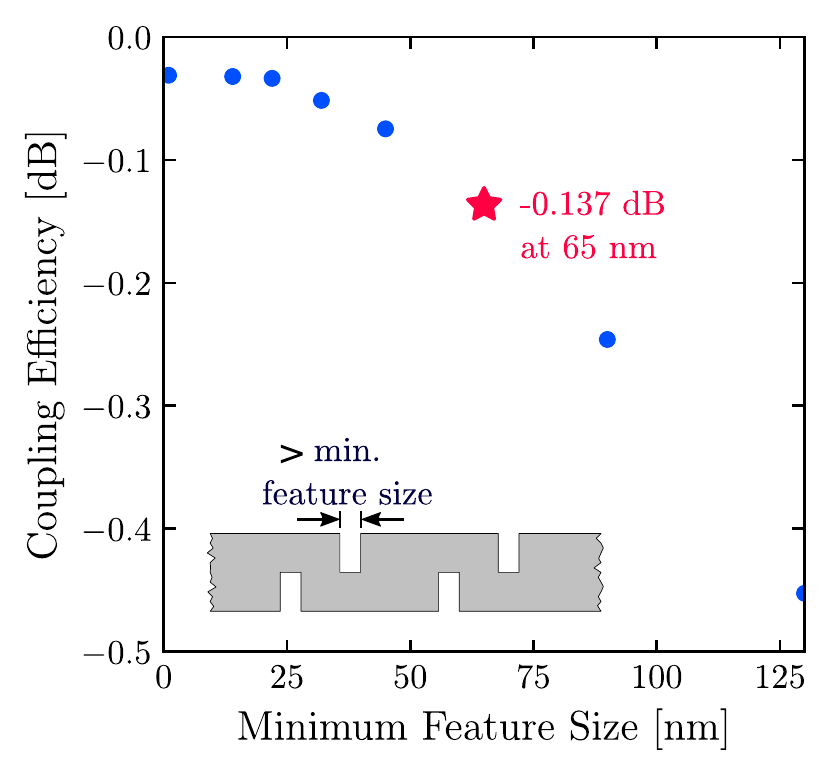}
		\caption{Plot of the coupling efficiency for perfectly vertical grating couplers optimized with a minimum feature size constraint. Using the ``ideal'' optimized result as a starting point, additional constrained optimizations are performed in order to design gratings that are fabricable with lithography that has a limited resolution.  Optimizations are performed for minimum feature sizes that roughly correspond to existing and future lithography capabilities.}
		\label{fig:fom_vs_feature}
\end{figure}

For the purpose of exploring the impact of minimum feature size on grating coupler efficiency, we run a series of three optimizations for a specific minimum feature size.  Beginning with the optimized structure shown in Fig. \ref{fig:Silicon_2L_PV_results}, we introduce a penalty function which is very weakly weighted (i.e. the maximum value of the penalty function is less than one) and then optimize the structure using our modified figure of merit for 25 iterations.  We repeat this process two more times, using the previous result as the starting structure for the next optimization and increasing the weight and sharpness of the penalty function each time.  

The result of this process, which we performed for a set of minimum feature sizes roughly corresponding to current and future lithography technologies, is shown in Fig. \ref{fig:fom_vs_feature}.  Using constrained optimization, we are able to maintain exceptionally high efficiencies out to more practical feature sizes.  Of particular interest is the \SI{65}{\nano\meter} constraint which, due to the maturity of \SI{65}{\nano\meter} CMOS node, shows future promise for silicon photonics \cite{gunn_cmos_2006}.  For a minimum feature size of \SI{65}{\nano\meter}, we have achieved an optimized efficiency of 96.68\% (\SI{-0.137}{\decibel}).  Furthermore, better than \SI{-0.5}{\decibel} is achievable out to a minimum feature size of over \SI{130}{\nano\meter}, roughly corresponding to technology nodes that are already used in commercial nanophotonic settings \cite{mekis_grating-coupler-enabled_2011}.  As the minimum feature size is further increased, the efficiency of the optimized design begins to fall off quickly.  This is a direct consequence of our inability to match to the gradually increasing Gaussian mode; when features are required to be on the same order of magnitude as the period of the grating, we lose much of our ability to control the light output from the grating.

\begin{figure}[!ht]
		\centering
		\includegraphics[width=\columnwidth]{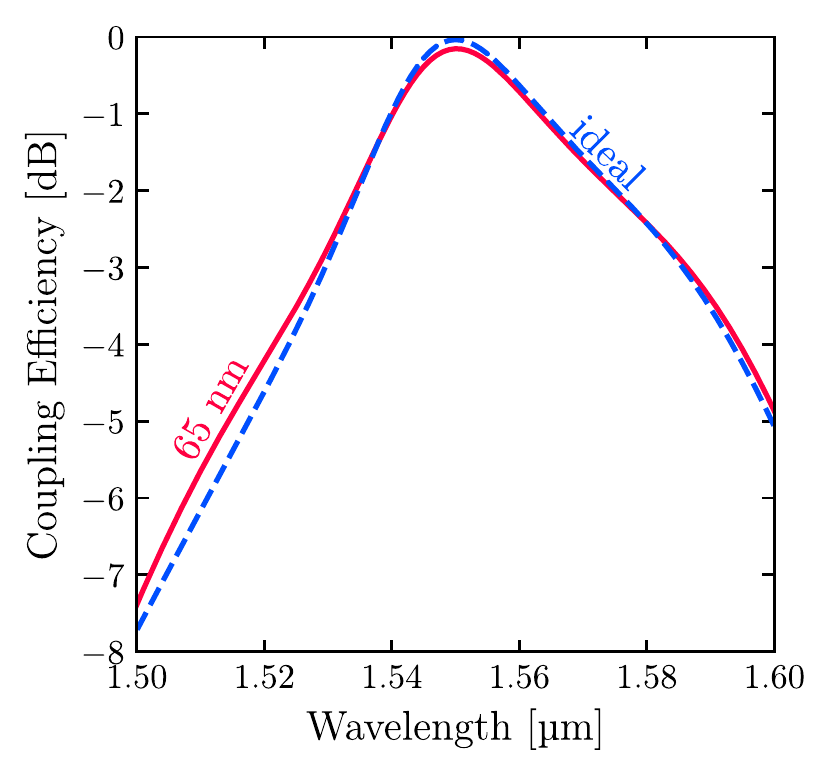}
		\caption{Plot of coupling efficiency as a function of wavelength for the ideal optimized result (blue dashed line) and the \SI{65}{\nano\meter} constrained minimum feature size result (red solid line).  Both cases achieve a high peak coupling efficiency at \SI{1550}{\nano\meter} as well as a modest \SI{1}{\decibel} bandwidth of \SI{24}{\nano\meter}.}
		\label{fig:fom_broadband}
\end{figure}

In addition to maintaining a high efficiency, our feature-size-constrained designs also maintain a reasonably large bandwidth.  Fig. \ref{fig:fom_broadband} shows the coupling efficiency plotted as a function of wavelength for the ``ideal'' unconstrained design and the \SI{65}{\nano\meter} design.  In both cases, the \SI{1}{\decibel} bandwidth is about \SI{24}{\nano\meter}.  This is to be expected since the bandwidth of a grating coupler is primarily dependent on the number of periods in the grating which is independent of the constrained feature size.  If necessary for a given application, this bandwidth could be increased by coupling into a fiber with a smaller mode field diameter which would allow us to reduce the number of periods within the grating.

\section{Conclusion}

Using efficient gradient-based shape optimization methods, we have design two-layer silicon grating couplers which couple \SI{1550}{\nano\meter} light perfectly vertically into single mode fibers with over 99\% efficiency.  Applying constrained optimization techniques, we have demonstrated that exceptionally high efficiencies can be maintained even when enforcing minimum feature sizes.  Of particular interest, we have achieved chip-to-fiber coupling efficiencies in excess of 96.9\% (\SI{-0.137}{\decibel}) that should be fabricable using \SI{65}{\nano\meter} lithography, a result which to our knowledge is the highest reported.

In our optimizations, we considered only 1D grating couplers (and hence simulations were performed in two dimensions).  In reality, the out of plane dimension places an important role in fiber coupling as well.  In particular, the transverse profile of the fundamental mode of the grating coupler does not exactly match the shape of the desired Gaussian mode.  As a consequence, real world efficiencies will be slightly lower than presented here (as is the case in much of the literature).  However, our optimization methods are sufficiently general that they can be applied to the full three dimensional problem in order to achieve three dimensional grating couplers which mitigate this additional source of mode mismatch.  We hope to tackle this challenge in future work.

In fact, there is a considerable amount of work left to be done in regards to optimizing full three dimensional grating couplers.  In particular, grating couplers which efficiently split polarized light is of considerable practical interest.  Such challenges are a perfect opportunity for us to make use of inverse electromagnetic design.  Between the rapid growth of silicon photonic processing and the power of these optimization tools, truly efficient vertical optical coupling is well on its way to becoming a reality.

\appendix
\section{Two-layer Grating Equations}
\label{sec:two_layer_equations}

\begin{figure}[h]
		\centering
		
		\includegraphics[width=\columnwidth]{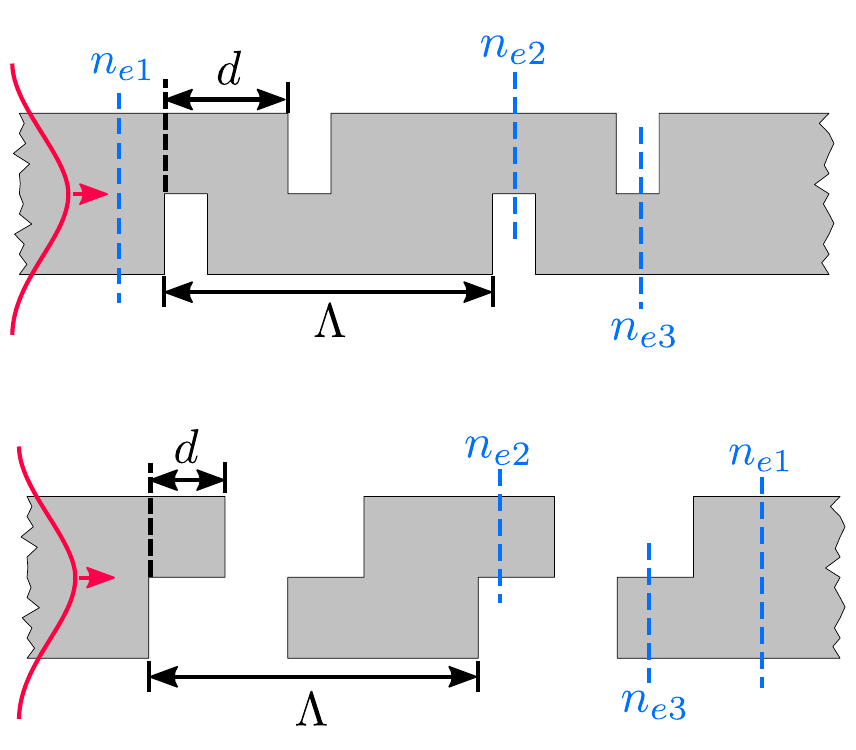}
		\caption{Diagrams of a two-layer grating with large duty factor (top) and smaller duty factor (bottom).  In both diagrams, $d$ denotes layer offset and $\Lambda$ denotes grating period (which for the sake of simplicity are assumed to be equal in both layers).  The lines marked $n_e$ represent the effective index in each section of the grating.  Although depicted here as being approximately equal, the thicknesses of the two layers need not be equal, and thus $n_{e2}$ and $n_{e3}$ will not necessarily be equal.}
		\label{fig:grating_parameter_diagram}
\end{figure}

In the design of two-layer gratings, it is important to understand how the grating parameters--effective index of the guided waves, period, and duty factor--affect the angle of the beam generated by the grating.  Knowing this relationship is essential from the standpoint of choosing an initial geometry for an optimization as well as understanding the chirp function that results from an optimization.

To derive a relationship relating the physical grating parameters to the angle of the generated beam, we begin with the grating equation which is true of any periodic grating coupler:

\begin{equation}
		k_g - \frac{2\pi m}{\Lambda} = k_0 \sin \theta \;\; .
		\label{eq:grating_equation}
\end{equation}

\noindent  In this expression, $k_g$ is the wavenumber of a guided wave propagating along the grating, $\Lambda$ is the grating period, $m$ is the diffraction order, $k_0$ is the wavenumber of the cladding material, and $\theta$ is the angle that the generated beam makes relative to the grating normal. For the sake of simplicity, $k_g$ is often approximated as being equal to the effective wavenumber of the mode propagating in the unetched waveguide at the input to the grating.  If we instead rewrite (\ref{eq:grating_equation}) in terms of the phase that a wave accrues over one period of the grating, we need not make any approximations for $k_g$.  Multiplying each side of Eq. (\ref{eq:grating_equation}) by $\Lambda$ and rearranging terms slightly yields,

\begin{equation}
		\phi_g = 2 \pi m + \Lambda k_0 \sin \theta
		\label{eq:phase_condition}
\end{equation}

\noindent where $\phi_g$ is the phase a guided wave in the grating must accrue over one period in order to generate a free-space wave propagating at an angle $\theta$ to normal. This expression defines the relationship between the physical parameters of the grating and the angle of the generated beam.

It is desirable to use Eq. (\ref{eq:phase_condition}) in order to derive expressions for the period of two-layer grating with a specified duty factor $D$ and desired coupling angle $\theta$.  To do so, there are three different grating configurations that must be considered independently. Fig. \ref{fig:grating_parameter_diagram} depicts the two configurations that we care about the most.  In the top diagram, the grating has a large duty factor (i.e. small gaps), which results in sets of fully separated gaps.  In this case, a wave propagating through the grating will travel through regions with up to three different effective refractive indices, which will determine the phase acquired by the wave along one period of the grating.  In the bottom diagram, the grating has a smaller duty factor which results in overlapping gap regions.  In these overlapping gap regions, the effective index is equal to that of the cladding index. A third possible configuration occurs when the duty factor becomes very small resulting in a grating which resembles an inverted version of the top diagram in Fig \ref{fig:grating_parameter_diagram}.  This is not a desirable regime of operation, however, and will not be considered.

In the case of a large duty factor, the phase acquired by a guided wave over one period of the grating is given by

\begin{equation}
		\phi_g = \Lambda\left[ (1-D) (k_{2} + k_{3}) + (2D - 1) k_{1} \right]
		\label{eq:phi_g_large_df}
\end{equation}

\noindent where $k_1$, $k_2$, and $k_3$ are the wavenumbers corresponding to the effective indices marked in Fig. \ref{fig:grating_parameter_diagram} and $D$ is the duty factor of the grating.  Substituting this result into (\ref{eq:phase_condition}) and solving for $\Lambda$ yields the grating period needed to generate a beam at the desired angle

\begin{align}
		\Lambda = m\lambda\biggl[&(1-D)(n_{e2}+n_{e3}) \nonumber\\&+ (2 D - 1) n_{e1} - n_{0}\sin\theta\biggr]^{-1}
		\label{eq:Lambda_large_df}
\end{align}

A second important quantity is the horizontal shift between the bottom and top layers of the grating, labeled $d$ in Fig. \ref{fig:grating_parameter_diagram}.  As discussed previously in this manuscript, the guided wave must accrue a phase of $\pi/2$ between the gap in the bottom layer and the gap in the top layer in order to achieve asymmetric emission from the grating.  Equating the phase that the wave picks up between the beginning of the first gap and the second gap to $\pi/2$ and solving for $d$ yields the shift between the layers in the case of a large duty factor:

\begin{equation}
		d = \frac{1}{n_{e1}} \left[ \frac{\lambda}{4} + \Lambda (1-D)(n_{e1} - n_{e2}) \right]
		\label{eq:layer_shift_large_df}
\end{equation}

Note that these two results are only valid when the gaps of the top and bottom grating do not overlap, that is when $d \ge (1-DF)\Lambda$.  Substituting for $\Lambda$ in this expression yields a definition for ``large duty factor:''

\begin{equation}
		D > \frac{(4 m - 1) n_{e2} - n_{e3} + n_0 \sin\theta + n_{e1}}{2 n_{e1} + (4 m - 1)n_{e2} - n_{e3}} \;\; .
		\label{eq:large_df_def}
\end{equation}

\noindent For a silicon grating clad in SiO$_2$ with roughly equal layer thicknesses, a ``large'' duty factor is over 74\%.

For duty factors which are below this value, we must use a different set of expressions corresponding to the case depicted in the bottom of Fig. \ref{fig:grating_parameter_diagram}.  The derivation of the period relevant to this configuration can be found in the same manner as was used for Equation (\ref{eq:Lambda_large_df}) and has as a result 

\begin{equation}
		\Lambda = \frac{m \lambda - \frac{\lambda}{4 n_{e2}}(n_{e2} + n_{e3} - n_{e1} - n_{0})}{D n_{e1} + n_{0} (1-D) - n_{0} \sin\theta}
		\label{eq:Lambda_small_df}
\end{equation}

\noindent where we have implicitly used the fact that the shift between layer needs to be 

\begin{equation}
		d = \frac{\lambda}{4 n_{e1}} \;\; .
		\label{eq:shift_small_df}
\end{equation}

\noindent Together, the expressions derived in this section provide us with the tools we need to design the most basic two-layer grating. 

\section*{Funding Information}
Office of Naval Research (ONR) grant numbers N00014-14-1-0505 and N00014-16-1-2237

 


\bibliography{references.bib}
%
\end{document}